
%
%
%
%
%
%
%
\baselineskip=1.4\baselineskip
\nopagenumbers
\footline{\hss \tenrm -- \folio\ -- \hss}
\hfuzz=10pt
\catcode`\@=11 

\def\nolabels{\def\wrlabel##1{}\def\eqlabel##1{}\def\reflabel##1{}}
\def\writelabels{\def\wrlabel##1{\leavevmode\vadjust{\rlap{\smash%
{\line{{\escapechar=` \hfill\rlap{\sevenrm\hskip.03in\string##1}}}}}}}%
\def\eqlabel##1{{\escapechar-1\rlap{\sevenrm\hskip.05in\string##1}}}%
\def\thlabel##1{{\escapechar-1\rlap{\sevenrm\hskip.05in\string##1}}}%
\def\reflabel##1{\noexpand\llap{\noexpand\sevenrm\string\string\string##1}}}
\nolabels
%
\global\newcount\secno \global\secno=0
\global\newcount\meqno \global\meqno=1
\global\newcount\mthno \global\mthno=1
\global\newcount\mexno \global\mexno=1
\global\newcount\mquno \global\mquno=1
\def\newsec#1{\global\advance\secno by1 
\global\subsecno=0\xdef\secsym{\the\secno.}\global\meqno=1\global\mthno=1
\global\mexno=1\global\mquno=1
\bigbreak\medskip\noindent{\bf\the\secno. #1}\writetoca{{\secsym} {#1}}
\par\nobreak\medskip\nobreak}
\xdef\secsym{}
\global\newcount\subsecno \global\subsecno=0
\def\subsec#1{\global\advance\subsecno by1
\bigbreak\noindent{\it\secsym\the\subsecno. #1}\writetoca{\string\quad
{\secsym\the\subsecno.} {#1}}\par\nobreak\medskip\nobreak}
\def\appendix#1#2{\global\meqno=1\global\mthno=1\global\mexno=1
\global\subsecno=0
\xdef\secsym{\hbox{#1.}}
\bigbreak\bigskip\noindent{\bf Appendix #1. #2}
\writetoca{Appendix {#1.} {#2}}\par\nobreak\medskip\nobreak}
%
%
\def\eqnn#1{\xdef #1{(\secsym\the\meqno)}\writedef{#1\leftbracket#1}%
\global\advance\meqno by1\wrlabel#1}
\def\eqna#1{\xdef #1##1{\hbox{$(\secsym\the\meqno##1)$}}
\writedef{#1\numbersign1\leftbracket#1{\numbersign1}}%
\global\advance\meqno by1\wrlabel{#1$\{\}$}}
\def\eqn#1#2{\xdef #1{(\secsym\the\meqno)}\writedef{#1\leftbracket#1}%
\global\advance\meqno by1$$#2\eqno#1\eqlabel#1$$}
%
%
\def\thm#1{\xdef #1{\secsym\the\mthno}\writedef{#1\leftbracket#1}%
\global\advance\mthno by1\wrlabel#1}
\def\exm#1{\xdef #1{\secsym\the\mexno}\writedef{#1\leftbracket#1}%
\global\advance\mexno by1\wrlabel#1}
%
\newskip\footskip\footskip14pt plus 1pt minus 1pt 
\def\f@@t{\baselineskip\footskip\bgroup\aftergroup\@foot\let\next}
\setbox\strutbox=\hbox{\vrule height9.5pt depth4.5pt width0pt}
\global\newcount\ftno \global\ftno=0
\def\foot{\global\advance\ftno by1\footnote{$^{\the\ftno}$}}
%
\newwrite\ftfile
\def\footend{\def\foot{\global\advance\ftno by1\chardef\wfile=\ftfile
$^{\the\ftno}$\ifnum\ftno=1\immediate\openout\ftfile=foots.tmp\fi%
\immediate\write\ftfile{\noexpand\smallskip%
\noexpand\item{f\the\ftno:\ }\pctsign}\findarg}%
\def\footatend{\vfill\eject\immediate\closeout\ftfile{\parindent=20pt
\centerline{\bf Footnotes}\nobreak\bigskip\input foots.tmp }}}
\def\footatend{}
%
%
\global\newcount\refno \global\refno=1
\newwrite\rfile
\def\ref{\the\refno\nref}
\def\bref{\nref}
\def\nref#1{\xdef#1{\the\refno}\writedef{#1\leftbracket#1}%
\ifnum\refno=1\immediate\openout\rfile=refs.tmp\fi
\global\advance\refno by1\chardef\wfile=\rfile\immediate
\write\rfile{\noexpand\item{[#1]\ }\reflabel{#1\hskip.31in}\pctsign}\findarg}
\def\findarg#1#{\begingroup\obeylines\newlinechar=`\^^M\pass@rg}
{\obeylines\gdef\pass@rg#1{\writ@line\relax #1^^M\hbox{}^^M}%
\gdef\writ@line#1^^M{\expandafter\toks0\expandafter{\striprel@x #1}%
\edef\next{\the\toks0}\ifx\next\em@rk\let\next=\endgroup\else\ifx\next\empty%
\else\immediate\write\wfile{\the\toks0}\fi\let\next=\writ@line\fi\next\relax}}
\def\striprel@x#1{} \def\em@rk{\hbox{}}

\def\addref#1{\immediate\write\rfile{\noexpand\item{}#1}} 
\def\startrefs#1{\immediate\openout\rfile=refs.tmp\refno=#1}
\def\xref{\expandafter\xr@f}\def\xr@f[#1]{#1}
\def\refs#1{[\r@fs #1{\hbox{}}]}
\def\r@fs#1{\edef\next{#1}\ifx\next\em@rk\def\next{}\else
\ifx\next#1\xref #1\else#1\fi\let\next=\r@fs\fi\next}
%

%
\newwrite\ffile\global\newcount\figno \global\figno=1
\def\fig{fig.~\the\figno\nfig}
\def\nfig#1{\xdef#1{fig.~\the\figno}%
\writedef{#1\leftbracket fig.\noexpand~\the\figno}%
\ifnum\figno=1\immediate\openout\ffile=figs.tmp\fi\chardef\wfile=\ffile%
\immediate\write\ffile{\noexpand\medskip\noexpand\item{Fig.\ \the\figno. }
\reflabel{#1\hskip.55in}\pctsign}\global\advance\figno by1\findarg}
\def\xfig{\expandafter\xf@g}\def\xf@g fig.\penalty\@M\ {}
\def\figs#1{figs.~\f@gs #1{\hbox{}}}
\def\f@gs#1{\edef\next{#1}\ifx\next\em@rk\def\next{}\else
\ifx\next#1\xfig #1\else#1\fi\let\next=\f@gs\fi\next}
\newwrite\lfile
{\escapechar-1\xdef\pctsign{\string\%}\xdef\leftbracket{\string\{}
\xdef\rightbracket{\string\}}\xdef\numbersign{\string\#}}

\def\writestop{\def\writestoppt{\immediate\write\lfile{\string\pageno%
\the\pageno\string\startrefs\leftbracket\the\refno\rightbracket%
\string\def\string\secsym\leftbracket\secsym\rightbracket%
\string\secno\the\secno\string\meqno\the\meqno}\immediate\closeout\lfile}}
\def\writestoppt{}\def\writedef#1{}
\def\seclab#1{\xdef #1{\the\secno}\writedef{#1\leftbracket#1}\wrlabel{#1=#1}}
\def\subseclab#1{\xdef #1{\secsym\the\subsecno}%
\writedef{#1\leftbracket#1}\wrlabel{#1=#1}}
\newwrite\tfile \def\writetoca#1{}
\def\leaderfill{\leaders\hbox to 1em{\hss.\hss}\hfill}
\def\writetoc{\immediate\openout\tfile=toc.tmp
   \def\writetoca##1{{\edef\next{\write\tfile{\noindent ##1
   \string\leaderfill {\noexpand\number\pageno} \par}}\next}}}

%
\catcode`\@=12 
%
%
%

\def\p{\partial}
\def\half{{\textstyle{1\over2}}}
%
%
\def\al{\alpha}
\def\be{\beta}
  
\def\de{\delta}  \def\De{\Delta}
\def\ep{\epsilon}

\def\th{\theta}

\def\la{\lambda}

\def\ph{\phi}  \def\Ph{\Phi}  

%
%

%

%
%

 \def\cF{{\cal F}}
\def\cH{{\cal H}}

\def\cO{{\cal O}}

 \def\cS{{\cal S}}

\def\ie{{\it i.e.\ }} \def\eg{{\it e.g.\ }}
%
\def\amsyes{y }

\ifx\answ\amsyes
\input amssym.def

\def\ZZ{{\Bbb Z}}

\def\bfg{{\frak g}}

\def\hg{{\widehat{\frak g}}}
\def\whg{\hg}
\def\sln{\frak{sl}_N}   \def\hsln{\widehat{\frak{sl}_N}}
\def\sltw{\frak{sl}_2}  \def\hsltw{\widehat{\frak{sl}_2}}

\def\son{\frak{so}_N}  \def\slkpe{\frak{sl}_{k+1}}
\else

\def\ZZ{{Z\!\!\!Z}}

\def\bfg{{\bf g}}

\def\hg{\widehat{\bf g}} \def\whg{\hg}

\def\sln{s\ell_N}   \def\hsln{\widehat{s\ell_N}}
\def\sltw{s\ell_2}  \def\hsltw{\widehat{s\ell_2}}

\def\son{so_N}
\def\slkpe{s\ell_{k+1}}
\fi
%
%
%
\def\AdM#1{Adv.\ Math.\ {\bf #1}}

\def\AnP#1{Ann.\ Phys.\ {\bf #1}}
\def\CMP#1{Comm.\ Math.\ Phys.\ {\bf #1}}

\def\IJMP#1{Int.\ J.\ Mod.\ Phys.\ {\bf #1}}
\def\InM#1{Inv.\ Math.\ {\bf #1}}

\def\JPA#1{J.\ Phys.\ {\bf A{#1}}}
\def\JRAM#1{J.\ reine angew.\ Math.\ {\bf {#1}}}
\def\JSP#1{J.\ Stat.\ Phys. {\bf {#1}}}
\def\LEM#1{L'Enseignement Math\'ematique {\bf {#1}}}
\def\LMP#1{Lett.\ Math.\ Phys.\ {\bf #1}}

\def\MPL#1{Mod.\ Phys.\ Lett.\ {\bf #1}}
\def\NPB#1{Nucl.\ Phys.\ {\bf B#1}}
\def\PLB#1{Phys.\ Lett.\ {\bf {#1}B}}
\def\PNAS#1{Proc.\ Natl.\ Acad.\ Sci. USA {\bf #1}}

\def\PRL#1{Phys.\ Rev.\ Lett.\ {\bf #1}}

\def\SMD#1{Sov.\ Math.\ Dokl.\ {\bf {#1}}}
\def\SPJETP#1{Sov.\ Phys.\ J.E.T.P.\ {\bf #1}}

%
%
%
\def\ni{\noindent}
\def\ch{{\rm ch}}
\def\qbin#1#2{\left[\matrix{ {#1} \cr {#2} \cr} \right]}
\def\vac{|0\rangle}
\def\CVO#1#2#3{\!\left( \matrix{ #1 \cr #2 \ #3 \cr} \right)\!}
\def\CVOS#1#2#3#4{\!\left( \matrix{ #1 \cr #2 \ #3 \cr} \right)_{\!#4}}
\def\min{{\rm min}}
%
%
\bref\AL{
I.~Affleck and A.W.W.~Ludwig,
\NPB{360} (1991) 641; \PRL{68} (1992) 1046;
A.W.W.~Ludwig and I.~Affleck, \NPB{428} (1994) 545.}

\bref\FLS{
P.~Fendley, A.W.W.~Ludwig and H.~Saleur, {\it Exact conductance
through point contacts in the $\nu=1/3$ fractional quantum Hall
effect}, \PRL{}(1995), {\it
in print}, {\tt (cond-mat/9408068)}.}

\bref\FSW{
P.~Fendley, H.~Saleur and N.~Warner, \NPB{430} (1994) 577,
{\tt (hep-th/9406125)}.}

\bref\KNS{
A.~Kuniba, T.~Nakanishi and J.~Suzuki,
\MPL{A8} (1993) 1649, {\tt (hep-th/9301018)}.}

\bref\KKMM{
R.~Kedem, T.~Klassen, B.~McCoy and E.~Melzer,
\PLB{304} (1993) 263, {\tt (hep-th/9211102)};
\PLB{307} (1993) 68, {\tt (hep-th/9301046)};
S.~Dasmahapatra, R.~Kedem, T.~Klassen, B.~McCoy and E.~Melzer,
\JSP{74} (1994) 239, {\tt (hep-th/9303013)};
R.~Kedem, B.~McCoy and E.~Melzer,
{\it The sums of Rogers, Schur and Ramanujan and the Bose-Fermi
correspondence in $1+1$-dimensional quantum field theory},
{\tt (hep-th/9304056)};
E.~Melzer, \LMP{31} (1994) 233,
{\tt (hep-th/9312043)}.}

\bref\Be{
A.~Berkovich, \NPB{431} (1994) 315, {\tt (hep-th/9403073)};
A.~Berkovich and B.~McCoy, {\it Continued fractions and fermionic
representations for characters of $M(p,p')$ minimal models},
{\tt (hep-th/9412030)}.}

\bref\Ki{
A.N.~Kirillov, {\it Dilogarithm identities}, {\tt (hep-th/9408113)}.}

\bref\FQ{
O.~Foda and Y.-H. Quano,
{\it Polynomial identities of the Rogers--Ramanujan type},
{\tt (hep-th/9407191)}; {\it Virasoro character identities from the
Andrews--Bailey construction}, {\tt (hep-th/9408086)}.}

\bref\LW{
J.~Lepowsky and R.L.~Wilson, \PNAS{78} (1981) 7254;
\AdM{45} (1982) 21; \InM{77} (1984) 199; \InM{79} (1985) 417.}

\bref\FNO{
B.~Feigin, T.~Nakanishi and H.~Ooguri, \IJMP{A7} Suppl.\ 1A
(1992) 217.}

\bref\FS{
B.~Feigin and A.~Stoyanovsky, {\it Quasi-particle models for the
representations of Lie algebras and geometry of flag manifold},
{\tt (hep-th/9308079)}.}

\bref\BPS{
D.~Bernard, V.~Pasquier and D.~Serban, \NPB{428} (1994) 612,
{\tt (hep-th/9404050)}.}

\bref\BLS{
P.~Bouwknegt, A.~Ludwig and K.~Schoutens,
\PLB{338} (1994) 448, {\tt (hep-th/9406020)}; {\it Affine
and Yangian Symmetries in $SU(2)_1$ Conformal Field Theory},
to appear in the proceedings of the 1994 Trieste Summer School on
``High Energy Physics and
Cosmology,'' Trieste, July 1994, {\tt (hep-th/9412199)}.}

\bref\Is{
S.~Iso, {\it Anyon basis of $c=1$ conformal field theory},
{\tt (hep-th/9411051)}.}

\bref\Ge{
G.~Georgiev,
{\it Combinatorial constructions of modules for infinite-dimensional
Lie algebras, I. Principal subspace}, {\tt (hep-th/9412054)}.}

\bref\HS{
F.D.M.~Haldane, \PRL{60} (1988) 635;
B.S.~Shastry, \PRL{60} (1988) 639.}

\bref\HHTBP{
F.D.M.~Haldane, Z.N.C.~Ha, J.C.~Talstra, D.~Bernard and
V.~Pasquier, \PRL{69} (1992) 2021.}

\bref\BGHP{
D.~Bernard, M.~Gaudin, F.D.M.~Haldane and V.~Pasquier,
\JPA{26} (1993) 5219, {\tt (hep-th/9301084)}.}

\bref\FR{
L.~Faddeev and N.~Reshetikhin, \AnP{167} (1986) 227;
N.~Reshetikhin, \JPA{24} (1991) 3299;
P.~Fendley, \PRL{71} (1993) 2485, {\tt (cond-mat/9304031)}.}

\bref\MS{
G.~Moore and N.~Seiberg, \CMP{123} (1989) 177;
{\it Lectures on RCFT},
in ``Superstrings '89,'' Proceedings Trieste 1989.}

\bref\DL{
C.~Dong and J.~Lepowsky, {\it Generalized vertex algebras and relative
vertex operators}, Prog.\ in Math.\ {\bf 112} (Birkh\"auser, Boston, 1993).}

\bref\ZF{
A.B.~Zamolodchikov and V.A.~Fateev,
\SPJETP{62} (1985) 215; \SPJETP{63} (1986) 913.}

\bref\Dr{
V.G.~Drinfel'd, \SMD{32} (1985) 254; \SMD{36} (1988) 212;
{\it Quantum Groups}, in
Proc.\ of the International Congress of Mathematicians (Berkeley, 1986).}

\bref\CP{
V.~Chari and A.~Pressley, \LEM{36} (1990) 267;
\JRAM{417} (1991) 87.}

\bref\AGS{
L.~Alvarez-Gaum\'e, C.~Gomez and G.~Sierra, \NPB{319} (1989) 155;
\NPB{330} (1990) 347.}

\bref\KP{
V.~Kac and D.~Peterson, \AdM{53} (1984) 125.}

\bref\Sc{
K.~Schoutens, \PLB{331} (1994) 335, {\tt (hep-th/9401154)}.}

%
%
\line{\hfil USC-94/20}
\line{\hfil UCSB-TH-94}
\line{\hfil PUPT-1522}
\line{\hfil {\tt hep-th/9412108}}
\bigskip

\leftline{{\bf SPINON BASIS FOR HIGHER LEVEL $SU(2)$ WZW MODELS}}
\vskip1cm

\leftline{Peter Bouwknegt$^1$, Andreas W.W.\  Ludwig$^2$ and
Kareljan Schoutens$^3$}
\vskip.3cm

\leftline{$^1$ Department of Physics and Astronomy, U.S.C., Los Angeles,
CA 90089-0484}
\leftline{$^2$ Department of Physics, University of California, Santa
Barbara, CA~93106}
\leftline{$^3$ Joseph Henry Laboratories, Princeton University,
Princeton, NJ~08544}
\vskip1cm

\ni {\bf Abstract:}
We propose a spinon basis for the integrable highest weight
modules of $\hsltw$ at levels $k\geq1$, and
discuss the underlying Yangian symmetry. Evaluating the characters
in this spinon basis provides new quasi-particle type expressions for
the characters of these integrable modules, and explicitly exhibits the
structure of an RSOS times a Yangian part, known \eg from $S$-matrix
results. We briefly discuss generalizations to other groups and
more general conformal field theories.

%
\newsec{Introduction}

Conventionally, the Hilbert space
of (rational) two dimensional conformal field theories (RCFT)
is described in terms of a chiral algebra that acts on a finite set of
fields that are primary with respect to this chiral algebra.
This procedure leads to so-called Verma-module bases of RCFT's,
and gives rise to `bosonic type' formulas for the characters
(\ie torus partition functions) of such conformal field theories,
reflecting this particular choice for the basis of the Hilbert space.
On the other hand, in a conformal field theory (CFT) there
are many bases of Hilbert space, none of which is a priori
distinguished. Which of those different bases is most
useful depends on the question that is being asked (see \eg
[\AL,\FLS]). This is quite different in a massive
field theory, where a basis of massive {\it  particles} is
naturally distinguished. In general, considering massive
perturbations of CFT's, and their corresponding particle
bases, one may generate, at least conceptually, ``quasi-particle''
bases of CFT's, by letting the mass tend to zero. This provides
a way to define the notion of a massless ``quasi-particle,''
which has recently  proven to be very useful
in cases where the massive perturbation, used to define
such a basis, is Yang-Baxter integrable [\FSW,\FLS].
The so-generated bases inherit a particle (``Fock-space'') like
structure, which is not manifest
in the Verma-module type basis of the {\it same} CFT.
Each basis of the Hilbert space
of a given CFT gives rise to a particular way of writing
the  partition function on a torus.
The equality of the different ways to write
the torus partition function, in different bases, gives rise
to remarkable identities. Moreover, bases of massless
quasiparticles appear to be deeply related to Yangian and affine
quantum symmetries, that are often present even when
conformal and scale invariance is broken. Therefore, we expect
that a better understanding of such bases of CFT's will also
provide valuable insights into perturbed CFT's.

Analysis of the thermodynamic Bethe Ansatz, arising
from (Yang-Baxter) integrable perturbations of CFT's, for a variety
of models, has recently led to a wealth of conjectures for so-called
quasi-particle (or fermionic) type characters for
conformal field theories (see, in particular [\KNS,\KKMM]),
some of which have been proven through
$q$-analysis (see \eg [\Be--\FQ]).
Most of these results still lack an
interpretation (and/or proof) in terms of a corresponding
structure of the Hilbert space of a
conformal field theory, \ie
a characterization  and/or construction of the corresponding ``quasi-particle''
basis of the Hilbert space (see however [\LW--\Ge]).

A particular interesting model illustrating the issues above
is the so-called Haldane-Shastry long-range spin chain [\HS], which is
integrable and has Yangian symmetry (even for finite chains).
Its low energy sector is identical to
a well-known conformal field
theory, namely the $SU(2)$ level-$1$ Wess-Zumino-Witten model [\HHTBP,\BGHP].
While the description of the Hilbert space
of the $SU(2)$ level-$1$ WZW model in terms of its chiral
algebra, \ie $\hsltw$, is complicated due to the existence
null vectors, it was found that
the Hilbert space has a very simple structure [\BPS,\BLS], originating
from the underlying Yangian symmetry: it
may be viewed as a ``Fock space'' of massless `spinon particles,' which
satisfy generalized commutation relations [\BLS] (and no other relations).
No similar description has been known for higher levels.
However, Bethe ansatz $S$-matrix calculations [\FR] suggest that the
higher level models might be described by  massless
`spinons' (similar to those at level-$1$) and `kinks' (reflecting
an RSOS structure).  In this paper we propose such a basis of the
Hilbert space of the higher level $SU(2)$ WZW conformal field theory
in terms of the modes
of the spin-$1/2$ affine primary chiral vertex operator,
along the lines of ref.\ [\BLS].
The RSOS structure (`kinks') is provided by the sequence of fusions
of the chiral vertex operator (which is absent at level-$1$),
and represents the ``non-abelian statistics'' of the chiral
vertex operators at level greater than one.
We also argue that there is an underlying Yangian symmetry.
Some justification for our proposal of the higher level spinon
basis comes from the resulting quasi-particle type
expressions for the characters, which
we have been numerically verified to high order.

The paper is organized as follows. In section 2
we will propose, along the lines of [\BLS], a basis for the spinon
Fock space at level $k$, and we will argue
that this space decomposes into a
direct sum of integrable highest weight modules
(corresponding to $SU(2)$ level-$k$ affine primaries)
-- each integrable
module occurring exactly once. In section 3 we will argue
that on the spinon
Fock space (and hence on the integrable modules) we can define the
action of the Yangian $Y(\sltw)$ under which the spinon Fock space
is fully reducible. In section 4 we will derive
new (quasi-particle type)
character formulas for the integrable highest weight modules of
$(\hsltw)_k$ and, finally, in section 5 we will
make some remarks on the results and discuss generalizations to other
groups and more general conformal field theories.

%
\newsec{Spinon basis for $(\widehat{\sltw})_k$ integrable highest
weight modules}

The affine Lie algebra $\hsltw$ at level-$k$ is defined by means of
the commutators
\eqn\eqBa{
[J_m^a, J_n^b] = f^{ab}{}_{c} J_{m+n}^c + k m \de_{m+n} d^{ab}\,,
}
where $m,n\in \ZZ$ and the adjoint index takes values $a= (++),3,(--)$.
The metric is determined by $d^{(++)(--)}=1,\, d^{33}=2$ and the
structure constants follow from $f^{(++)(--)}{}_3=1$.

The integrable highest weight modules of $(\hsltw)_k$ will be denoted
by $L_j$. Here, the spin $j$ can take values $j = 0,\half,1,\ldots,
{k\over2}$. For the chiral vertex operators (CVO's)
$L_{j_1}\to L_{j_2}$ transforming in
the irreducible $\sltw$ representation of spin $j_3$, we will use
the notation of [\MS],
\ie
$\Ph\CVO{j_3}{j_2}{j_1}(z)$. They are
nonvanishing iff $j_2$ occurs in the fusion rule
$j_1\times j_3$, \ie $j_2 \in \{ |j_1-j_3|, \ldots, {\rm min}(j_1+j_3,
k-(j_1+j_3) )\}$. The spinons correspond to $j_3=\half$, and transform
according to
\eqn\eqBb{
[J_m^a,  \ph^\al\CVO{{1\over2}}{j_2}{j_1}(z)] =
  z^m \, (t^a\cdot \ph)^\al\CVO{{1\over2}}{j_2}{j_1}(z) \equiv
  z^m\, (t^a)^\al{}_\be\ph^\be\CVO{{1\over2}}{j_2}{j_1}(z) \,,
}
where $(t^{(++)})^-{}_+ = (t^{(--)})^+{}_- =1,\,
(t^3)^\pm{}_\pm = \pm1$.

The basic relations for the CVO's are the so-called braiding relations
(see \eg [\MS])
\eqn\eqBe{
\Ph\CVO{k_1}{j_1}{p}(z_1) \Ph\CVO{k_2}{p}{j_2}(z_2) = \sum_{p'}
B_{pp'}\! \left[ \matrix{ k_1 & k_2 \cr j_1 & j_2 \cr} \right]\!
\Ph\CVO{k_2}{j_1}{p'}(z_2) \Ph\CVO{k_1}{p'}{j_2}(z_1)\,,
}
and fusion relations
\eqn\eqBf{
\Ph\CVO{k_1}{j_1}{p}(z_1) \Ph\CVO{k_2}{p}{j_2}(z_2) =
\sum_{p'} F_{pp'}\!\left[ \matrix{ k_1 & k_2 \cr j_1 & j_2 \cr} \right]\!
\sum_{Q\in L_{p'}} \Ph^Q\CVO{p'}{j_1}{j_2}(z_2)
\langle Q| \Ph\CVO{k_1}{p'}{k_2}(z_1-z_2) | k_2\rangle \,,
}
where $Q\in L_{p'}$ denotes primary and descendant states
in the module  $L_{p'}$.  The braiding and fusion matrices,
$B_{pp'}\! \left[ \matrix{ k_1 & k_2 \cr j_1 & j_2 \cr} \right]\!$
and $ F_{pp'} \!\left[ \matrix{ k_1 & k_2 \cr j_1 & j_2 \cr} \right]\!$,
satisfy the pentagon and hexagon identities[\MS].

The CVO's $\Ph\CVO{j_3}{j_2}{j_1}(z)$ have conformal dimension
$\De(j_3)$ where
\eqn\eqBg{
\De(j) \equiv {j(j+1) \over k+2} \,,
}
and their mode expansion is given by
\eqn\eqBh{
\Ph\CVO{j_3}{j_2}{j_1}(z) = \sum_{n\in \ZZ}
\Ph\CVOS{j_3}{j_2}{j_1}{-n-(\De(j_2)-\De(j_1))} z^{n + (\De(j_2) -\De(j_1)
-\De(j_3))} \,.
}
In general, due to the non-locality of the CVO's, the modes will not
satisfy simple relations.  If, however, for a given $p$, the
braiding and fusion matrices $B_{pp'}$ and $F_{pp'}$ are nonvanishing
for one choice of $p'$ only, the equations \eqBe\ and \eqBf\ can be
combined into so-called generalized commutation relations.
For example,
\eqn\eqBi{
\sum_{l\geq0} C^{(2\De-1)}_l \left(
  \phi^\al\CVOS{{1\over2}}{0}{{1\over2}}{-m-l+\De}
  \phi^\be\CVOS{{1\over2}}{{1\over2}}{0}{-n+l-\De} -
  \left( \matrix{ m \leftrightarrow n+1
   \cr \al \leftrightarrow \be \cr} \right)
  \right) =  \ep^{\al\be} \de_{m+n,0} \,,
}
where $C^{(\al)}_l$ is defined by
\eqn\eqBj{
(1-x)^\al = \sum_{l\geq0} C^{(\al)}_l x^l\,,
}
and $\De = \De({{1\over2}})$.
Algebras of the type \eqBi\ are known as `generalized vertex algebras' [\DL].
Other examples include the so-called parafermion algebras [\ZF] or
$Z$-algebras [\LW].
Unfortunately, when the braiding and/or fusion matrix is not
one-dimensional (\ie the case of `non-abelian statistics'),
we do not know how to translate the content of
equations \eqBe\ and \eqBf\ into generalized commutation relations
between the modes.

Let $\cF$ denote the spinon Fock space, \ie the collection of
states of the form
\eqn\eqBk{
\ph^{\al_N}\CVOS{{1\over2}}{j_N}{j_{N-1}}{-\De_{N}-n_N} \cdots\
\ph^{\al_2}\CVOS{{1\over2}}{j_2}{j_1}{-\De_{2}-n_2}
\ph^{\al_1}\CVOS{{1\over2}}{j_1}{0}{-\De_{1}-n_1} \vac \,,
}
with $n_i + \De_i\geq0$, and where
the spins $\{ j_1,\ldots,j_N\}$ run over the set of spins allowed
by the fusion rules. Also we have put $\De_k = \De(j_k) - \De(j_{k-1})$.

Clearly, because of the relations \eqBe\ and \eqBf,
the basis \eqBk\ is overcomplete.  In sub-channels $0\to {1\over2}
\to 0$ we can `straighten' the basis by means
of the generalized commutation relation \eqBi.
Furthermore, by empirically
matching the low lying energy states for low levels $k$ against the
known irreducible characters for arbitrary fusion
channels $(...,j_3,j_2,j_1,0)$,  we found evidence that
a set of (independent) basis vectors is provided by the states
\eqn\eqBl{ \eqalign{
& \ph^-\CVOS{{1\over2}}{j_{M+N}}{j_{M+N-1}}{-\De_{M+N}-n_{M+N}} \cdots\
\ph^-\CVOS{{1\over2}}{j_{M+1}}{j_{M}}{-\De_{M+1}-n_{M+1}} \cr
& \qquad\times \ph^+\CVOS{{1\over2}}{j_{M}}{j_{M-1}}{-\De_{M}-n_M}\cdots
\ \ph^+\CVOS{{1\over2}}{j_{1}}{0}{-\De_{1}-n_1} \vac \,,\cr}
}
where the modes $n_i\equiv n_{i,\min} + \tilde{n}_i$ satisfy
$\tilde{n}_M\geq \tilde{n}_{M-1}\geq \ldots \geq \tilde{n}_1 \geq 0$,
$\tilde{n}_{M+N}\geq \tilde{n}_{M+N-1}\geq \ldots \geq \tilde{n}_{M+1} \geq 0$,
and where $n_{1,\min},\ldots,n_{M+N,\min}$ is a `minimal allowed
mode sequence' corresponding to the given fusion channel (Bratteli
diagram) constructed as follows
\eqn\eqBLa{ \eqalign{
n_{1,\min} & =0\,,\cr
n_{i+1,\min} & = \cases{ n_{i,\min}+1 & if $j_{i+1}=j_{i-1}<j_i$ \cr
                       n_{i,\min}   & otherwise $  \,.$\cr}\cr}
}
The rule \eqBLa\ is consistent with \eqBi, but at this point we do not
know how to derive \eqBLa\ from \eqBe\ and \eqBf.  Clearly, this
deserves further investigation.

The spinon Fock space $\cF$ is, in fact, an $(\widehat{\sltw})_k$ module
but,
a priori, this module does not have to be integrable (or even a direct
sum of integrable modules). However, in this particular case, we claim
that we do have $\cF \cong \bigoplus_{j=0}^{k/2} L_j$.
To prove that $\cF$ is indeed a
direct sum of integrable modules one would have to show
that the operator $S^{(k+1)}_N = \sum_{i_1+\ldots i_{k+1}=N}
J^{(++)}_{i_1} \cdots J^{(++)}_{i_{k+1}}$ annihilates all states
in $\cF$ for all $N\in \ZZ$ [\FS].
The highest weight vectors $|j\rangle$ of $L_j$ are then given by
\eqn\eqBm{
|j\rangle = \ph^+\CVOS{{1\over2}}{j}{j-{1\over2}}{-\De(j)
  +\De(j-{1\over2})}\cdots\
  \ph^+\CVOS{{1\over2}}{{1\over2}}{0}{-\De({1\over2})} \vac \,.
}
Clearly, these are $\sltw$ highest weight vectors. That they are in fact
$\hsltw$ highest weight vectors follows from
\eqn\eqBn{
( J^{(++)}_{-1} )^{N} |j\rangle = 0\,,
}
for all $N\geq k+1 - 2j$. Assuming the correctness of the basis
\eqBl, \eqBLa, equation \eqBn\ is a
simple consequence of the fact that the basis in $\cF$ does
not contain any states at that particular $+$-spinon number and
energy eigenvalue $L_0$.

To summarize this section, we conjecture
that the spinon Fock space $\cF$ has a basis given by
\eqBl\ and \eqBLa, and
is precisely the sum of all integrable highest weight modules $L_j$
of $(\hsltw)_k$. For level $k=1$ this conjecture has been proven in [\BLS].
In section 4 we will, assuming the validity of the conjecture, compute a
quasi-particle type expression for the character of $L_j$.  The correctness
of these characters, verified numerically for low level, isospin and
energy, strongly supports the validity of the conjecture.

%
\newsec{Yangian and Hecke symmetry}

We recall that the level-1 $\hsltw$ integrable modules carry
a (fully reducible) representation of the Yangian $Y(\sltw)$
[\HHTBP,\BPS,\BLS].
The Yangian generators were suggested by taking the low energy
limit of an XXX spin chain model with long-range interactions, the
so-called Haldane-Shastry model [\HS]. This spin chain
has exact Yangian symmetry and
 goes to the level-1 $\hsltw$ conformal field theory in the low energy limit,
as for the usual nearest neighbour
Heisenberg XXX chain.
Explicitly, we have [\HHTBP]
\eqn\eqCa{
Q_0^a = J_0^a\,,\qquad\qquad Q_1^a = \half f^a{}_{bc} \sum_{m>0}
  J_{-m}^b J_m^c\,.
}
To prove that \eqCa\ satisfy the relations of $Y(\sltw)$ [\Dr]
is a straightforward, albeit tedious, excercise.
In practice, one finds that the Yangian relations
are satisfied modulo fields which are null if and only if $k=1$.
On the other hand, there are results (in particular those of [\FR])
suggesting
that the Yangian symmetry {\it does} generalize to higher level.

It proves to be convenient to work
out the action of \eqCa\ on the spinon basis.
Moreover, to describe the action on the spinon basis it is convenient to work
with generating series for the $N$-spinon basis elements, \ie
\eqn\eqCaa{
\ph^{\al_N}\CVO{{1\over2}}{j_N}{j_{N-1}}(w_N )
  \cdots\  \ph^{\al_1}\CVO{{1\over2}}{j_1}{0}(w_1) \vac \,\in\,
  \cF(\!(w_1,\ldots,w_N)\!) \,.
}
On $\cF(\!(w_1,\ldots,w_N)\!)$, the action of \eqCa\ is given
by the following differential operators [\BPS]
\eqn\eqCb{ \eqalign{
Q_0^a & = \sum_i t_i^a \,,\cr
Q_1^a & = -2 \sum_i D_i t_i^a + \half f^a{}_{bc}
 \sum_{i\neq j} \th_{ij} t_i^b t_j^c\,,\cr}
}
where
\eqn\eqCba{
D_i = w_i \p_{w_i}\,,\qquad\qquad  \th_{ij} = {w_i \over w_i-w_j} \,.
}
In fact, recognizing \eqCb\ as the Yangian generators of the
$\sltw$ Calogero-Sutherland-Moser
model for coupling value $\la = -\half$
(see \eg [\BGHP]) gives an easy proof
of the fact that, for $k=1$, the generators \eqCa\ satisfy the relations
of $Y(\sltw)$ [\BPS].

The analogous expression for $k>1$ on the 2-spinon states
\eqn\eqCca{
\ph^{\al_2}\CVO{{1\over2}}{0}{{1\over2}}(w_2 )
  \,\ph^{\al_1}\CVO{{1\over2}}{{1\over2}}{0}(w_1) \vac \,,
}
is again given by \eqCb, but now with
\eqn\eqCc{
D_i = w_i\p_{w_i} - {\textstyle{1\over4}}
  (1-4\De) \sum_{j\neq i} (\th_{ij}- \th_{ji})\,,
}
where $\De \equiv \De(\half) = 3/ 4(k+2)$.
Note, in particular, that for $k>1$ it is impossible to express
\eqCb\ in terms of affine currents.

The rationale behind \eqCb\ is as follows. The one-spinon states transform
in an irreducible ($2$-dimensional) representation of $\sltw$. This
representation can be extended to a representation of $Y(\sltw)$, the
so-called `evaluation representation' [\Dr,\CP] (the evaluation parameter is
given by $2n$, where $n$ is the mode of the spinon). This gives equation
\eqCb\ for $N=1$ (one spinon).
The action of $Y(\sltw)$ on multi-spinon states is,
in principle, given by a `co-multiplication.' The multi-spinon Fock space
is, however, not a direct product of one-spinon Fock spaces but is
constrained by the generalized commutation relations \eqBi\ as well as
the constraints on the allowed fusion channels. The correct co-multiplication
is the one that descends to the multi-spinon Fock space, \ie leaves
the generalized commutation relations invariant. It is straightforward
to show that \eqCb\ satisfies this property on the 2-spinon states \eqCca.
In fact, the improvement
term in $D_i$ \eqCc\ simply corresponds to a `gauge transformation' [\BGHP]
that effectively transforms the generalized commutation relations
\eqBi\ at level $k$ into the ones for $k=1$. Since the gauge transformation
does not affect the commutation relations $[D_i,D_j] = 0,\, [D_i,w_j]=
\de_{ij} w_i$,
the generators \eqCb\ again satisfy the relations of $Y(\sltw)$.

The commutant of $Y(\sltw)$ on integrable
$(\hsltw)_k$ modules for
level $k=1$ contains an infinite set of mutually commuting
`Hamiltonians.' The first two are explicitly given by
\eqn\eqCd{ \eqalign{
H_1 & = L_0 = \sum_i \left( D_i + \De \right) \,,\cr
H_2 & = 2 \sum_i \left( (D_i)^2 + 2\De D_i \right) -
  d_{ab} \sum_{i\neq j}\th_{ij}\th_{ji} t_i^a t_j^b\,.\cr}
}
where $D_i$ and $\th_{ij}$ are as in \eqCba. Note that (for $k=1$) the
Calogero-Sutherland-Moser
Hamiltonian $H_2$ can also be expressed as (see [\BPS])
\eqn\eqCe{
H_2 = d_{ab} \sum_{m>0} m J_{-m}^a J_{m}^b \,.
}
For $k>1$ it is easily seen that on the 2-spinon states \eqCca\
the Hamiltonians \eqCd, but now with $D_i$ given by \eqCc,
are again in the commutant of $Y(\sltw)$.

Although we have not been able to prove this,
we expect the Yangian symmetry at $k>1$ to generalize to multi-spinon
states.  We are lacking explicit expressions for the Yangian
generators on $N$-spinon states with $N>2$ (expressions of the
form \eqCb\ will not do), but we have observed that
the decomposition of $\cF$ into irreducible representations of $Y(\sltw)$
seems to be completely analogous to the one for level-$1$, explained in [\BLS].
In particular, it seems that
the highest weight vectors of $Y(\sltw)$ are given by the fully
polarized states, \ie containing only $\phi^+$ fields. For generic mode
sequences $n_1,\ldots,n_M$, the corresponding (irreducible)
Yangian representation is $2^M$ dimensional and, as an $\sltw$
representation, is isomorphic to the $M$-fold tensorproduct of the
spin-$\half$ representation. For sequences $n_1,\ldots,n_M$
containing pairs $(n_{i},n_{i+1})$ with the minimal possible increment
as allowed by the rules below \eqBl, the corresponding (irreducible)
Yangian representation is smaller, namely, one has to project
onto the symmetric part (\eg the triplet for $M=2$) of the corresponding
tensor product of doublet representations.

A new feature, as compared to $k=1$, is that in addition to the
Yangian we have the action of a Hecke algebra on the $N$-spinon
Fock space. Its
generators essentially correspond to the exchange of spinons in
\eqCaa\ by means of the braiding matrices.
Clearly, this defines a representation of the Braid group
${\cal B}_N$ but, as shown in \eg [\AGS], this
representation actually
factors through the Hecke algebra $\cH_N(q)$ for $q = \exp( 2\pi i/(k+2))$.

%
\newsec{Character formulas}

Given the spinon basis \eqBl\ the computation of the character
$\ch_{L_j}(z;q) = {\rm Tr\,}_{L_j} ( q^{L_0} z^{J_0^3} )$
is now straightforward. Consider the states with $(M,N)$ number
of $(+,-)$-spinons. The sum over the
spinon modes $\tilde{n}_1,\ldots,\tilde{n}_{M+N}$
contributes the usual factor
\eqn\eqDa{
\cS_{M,N}(z,q) = {1 \over (q)_M (q)_N }
z^{{1\over2} (M-N)}\,,
}
which, in terms of the total spinon number $m_1= M+N$ and isospin
$j' = {1\over2} (M-N)$, can also be written as
\eqn\eqDb{
\cS_{M,N}(z,q)\,=\, { 1  \over
  (q)_{{1\over2} m_1 + j'}  (q)_{{1\over2} m_1 - j'} } z^{j'}\, =\,
  {1\over (q)_{m_1}} \qbin{m_1}{{1\over2} m_1 - j'}
  z^{j'} \,\equiv\, \cS^{j'}_{m_1}(z;q) \,.
}
Here we have introduced, as usual, the $q$-numbers and $q$-binomial
by
\eqn\eqDba{
(q)_N = \prod_{k=1}^N (1-q^k) \,,\qquad\qquad
  \qbin{M}{N} = { (q)_M \over (q)_N (q)_{M-N} }\,.
}
The combinatorics involved in performing the sum over fusion channels
of length $m_1=M+N$,
with the minimal mode sequence $n_{1,\min},\ldots,n_{M+N,\min}$, and
such that $j_{M+N}=j$ is formally similar to the
combinatorics that we used in [\BLS] to compute
spinon contributions to the Virasoro characters.
The only difference, as compared to [\BLS],
is that we are now dealing
with a truncated Bratteli diagram (as dictated by the fusion rules).
This reflects itself in the fact that now there only
exists a {\it finite} set $\{m_2,\ldots,m_k\}$ labelling the
`ghost excitations.' [Here, $k$ is the level of $(\hsltw)_k$.]
Provided ${m_1\over2} - j \in \ZZ_{\geq0}$,
we find the following contribution
\eqn\eqDc{
q^{-{j\over 2} - {1\over4}m_1^2}
  \sum_{m_2,\ldots,m_k} q^{{1\over2} (m_1^2 + m_2^2
  + \cdots + m_k^2 - m_1m_2 - m_2m_3 -\cdots -m_{k-1}m_{k})}
  \prod_{i=2}^k \qbin{ {1\over2} (m_{i-1} + m_{i+1} + \de_{i,2j+1})}{m_i}\,,
}
where the summation is over all odd positive integers for
$m_{2j},m_{2j-2},m_{2j-4},\ldots$ and over the even positive integers
for the remaining ones (we set $m_{k+1}\equiv0$). Recall
that $m_1$ (not summed) is the total spinon number.
For $q=1$ the expression \eqDc\ gives the number of spin-$j$ representations
in the $m_1$-fold (truncated) tensor product of the spin-$\half$
representation of $U_q(\sltw)$ at $q = \exp( 2\pi i/(k+2))$.

Let us define for an arbitrary (symmetric) $k\times k$-matrix $K$ and
$k$-vector $u$, the $q$-series
\eqn\eqDd{
\Phi^{m_1}_K(u;q) = \sum_{m_2,m_3,\ldots,m_k}
  q^{ {1\over4} m\cdot K\cdot m} \prod_{i\geq2}
  \qbin{ {1\over2} ( (2-K) \cdot m + u )_i}{m_i} \,,
}
then \eqDc\ can be written as
\eqn\eqDe{
q^{-{j\over 2} - {1\over4}m_1^2} \,\Phi^{m_1}_{A_k}(u_j;q) \,,
}
where $A_k$ is the Cartan matrix of the Lie algebra $A_k \cong \slkpe$
and $u_j$ is the unit vector $(u_j)_i = \de_{i,2j+1}$.

Combining the ingredients, we have obtained the following expression
for the characters of the $(\hsltw)_k$ integrable modules
$L_j$ ($j=0,\half,\ldots,{k\over2}$)
\eqn\eqDf{
\ch_{L_j}(z,q) =  q^{ \De(j) - j/2 }
\sum_{M,N\geq0} q^{-{1\over4} (M+N)^2} \Ph^{M+N}_{A_k}(u_j;q)
\cS_{M,N} (z;q)\,.
}
For $k=1$ this reproduces the result of [\KKMM,\BPS,\BLS]. For $k>1$ this
quasi-particle form of the character, as well as the expressions
for the string functions [\KP]  $c_{2j'}^{2j}(q)$ that can immediately be
read off using \eqDb, appear to be new.
For low level and isospin, we have verified numerically
(typically up to $\cO(q^{25})$) that
these string functions do indeed coincide with the ones in \eg [\KP].
The correctness of the characters \eqDf\ strongly supports the
claims made in sections 2 and 3.

Similarly, expressions for the generating series $\Psi_{j,j'}(q)$ of
the number of $\sltw$ representations
of spin $j'$ in $L_j$ can be obtained from \eqDf,
\eqDb\ and the identity
\eqn\eqDg{
\qbin{m_1}{ {m_1\over2} - j'} - \qbin{m_1}{ {m_1\over2} - j' - 1 }  =
q^{-{m_1^2\over4} - {j'\over2}}\, \Phi_{A_\infty}^{m_1}(u_{j'};q)\,,
}
which follows from the analysis in [\BLS]. We find
\eqn\eqDh{
\Psi_{j,j'}(q) = q^{\De(j)-{j+j'\over 2}}\sum_{m_1}
  \Phi_{A_k}^{m_1}(u_j;q)\, {q^{-{1\over2} m_1^2} \over (q)_{m_1} }\,
  \Phi_{A_\infty}^{m_1}(u_{j'};q)\,.
}
This generalizes the result for $k=1$ (see eqn.\ (35) in [\BLS]) in which
case \eqDh\ corresponds to the irreducible Virasoro character
${\rm ch}_{h=j'{}^2}^{\rm Vir}(q)$, where
$j=0,\half$ for $j'$ integer or halfinteger, respectively.
Note that \eqDh\ coincides
with the $l \to \infty$ limit of the conjectured
branching functions for the coset $(\hsltw)_k \oplus (\hsltw)_l
/ (\hsltw)_{k+l}$ in [\KKMM] (for $j=j'=0$), as it should.

%
\newsec{Concluding remarks}

In this paper we have proposed a generalization
of the results of [\BLS] for
level-1 $\hsltw$ to higher level. We have seen that both
the spinon description of the integrable modules as well as
the Yangian symmetry are likely to
pertain at higher levels. In addition to
the Yangian symmetry, we have seen the occurrence of a Hecke
symmetry $\cH_N(q= \exp( 2\pi i/k+2))$
on the $N$-spinon subspace.
Using the proposed spinon
description of the spectrum we have derived new character formulas
for the level-$k$ integrable modules.

{}From the characters \eqDf\  it can be seen that the $N$-spinon subspace
naturally factorizes into a `Yangian part,' corresponding to $\cS_{M,N}(z;q)$,
and a `Hecke (or RSOS)
part,' corresponding to $\Phi^{m_1}_{A_k}(u_j;q)$. A similar factorization
was observed in the $S$-matrix for higher level $SU(2)$ WZW-models,
as well as in the analysis of the TBA for the generalized
(\ie higher $\sltw$ representations) integrable  XXX spin chain
models which, in the low energy limit, are known to give rise
to higher level $SU(2)$ WZW-models [\FR]. The spinon basis
in this paper gives a natural
explanation of these aforementioned results.
An interesting problem that immediately comes to mind is whether
there exists a corresponding generalization of the Haldane-Shastry model, \ie
a spin chain model (with long-range interactions) that has
exact Yangian symmetry and gives rise to higher level $\hsltw$ in the
low energy limit. A natural candidate, obviously, would be to
extend the usual Haldane-Shastry model with additional `local height
variables.'

Generalization of the results in this paper to $(\hsln)_k$
is, in principle, straightforward.
Again we expect to find that
the Fock space of the primary field in the vector representation of
$\sln$ decomposes into a direct sum of integrable representations --
each integrable representation occurring exactly once -- and that
the integrable representations carry a (fully reducible) representation
of $Y(\sln)$ (see [\Sc] for level $k=1$), whose action on
the spinon basis is related to the $\sln$
Calogero-Sutherland-Moser Yangian.

In general, \ie for other Lie algebras $\bfg$ and/or other representations
of $\bfg$ as well as for more general
conformal field theories such as coset models,
the situation is more complicated. For instance, for groups $\bfg$ other
than $\sln$ the integrable highest weight modules of $\whg$ do
not carry a representation of $Y(\bfg)$, simply because
finite-dimensional irreducible representations of $\bfg$ do, in general, not
extend to representations of $Y(\bfg)$. Another, more
important, difference is that the Fock spaces
corresponding to some chiral primary field in the theory do, in general,
not decompose into a direct sum of irreducible (integrable) modules.
Rather, in terms of quasi-particle language, one would say that
the corresponding conformal field theory corresponds to a
theory of quasi-particles with other than purely statistical
interactions.

That these two issues are sometimes closely related can be seen in
the following example. Consider $\bfg \cong \son$. The
spinor representation of $\son$ extends to a representation
of $Y(\son)$ [\Dr]. This then defines an action of $Y(\son)$
on the one-spinon Fock space of the spinon field of
$(\widehat{\son})_k$, which can
presumably be extended to an action on the
entire spinon Fock space by co-multiplication (as outlined in section 3).
However, for $N\neq 3$,
this Fock space is {\it bigger} than just a direct sum of
integrable modules. Upon projection to the integrable modules this
Yangian symmetry gets lost.\bigskip

%
%
\ni{\bf Acknowledgements:}  K.S. thanks the Physics Department of
UC Santa Barbara, where part of this work was done,
for hospitality. The research of K.S.\ was supported in
part by the National Science Foundation under grant
PHY90-21984. A.W.W.L.\ is a  Fellow of the A.P.\ Sloan Foundation.
\footatend\immediate\closeout\rfile\writestoppt
\baselineskip=14pt{\bigskip\noindent {\bf  References}}%
\bigskip{\frenchspacing%
\parindent=20pt\escapechar=` \input refs.tmp\vfill\eject}\nonfrenchspacing
\vfil\eject\end